\documentstyle[epsfig]{aipproc}
\begin{document}
\title{One-dimensional transport in bundles of single-walled carbon nanotubes}
\author{David H. Cobden$^\ast$, Jesper Nyg\aa rd$^\ast$, Marc Bockrath$^\dagger$\\ and Paul L. McEuen$^\dagger$}
\address{$^\ast$ \O rsted Laboratory, Niels Bohr Institute,\\ Universitetsparken 5, DK-2100 Copenhagen, Denmark\\
$^\dagger$ Department of Physics, University of California at Berkeley \\and
Lawrence Berkeley National Laboratory, Berkeley, CA 94720, USA}
\maketitle
\begin{abstract}
We report measurements of the temperature and gate voltage dependence for individual bundles (ropes) of single-walled nanotubes.  When the conductance is less than about $e^2/h$ at room temperature, it is found to decrease as an approximate power law of temperature down to the region where Coulomb blockade sets in.  The power-law exponents are consistent with those expected for electron tunneling into a Luttinger liquid.  When the conductance is greater than $e^2/h$ at room temperature, it changes much more slowly at high temperatures, but eventually develops very large fluctuations as a function of gate voltage when sufficiently cold.  We discuss the interpretation of these results in terms of transport through a Luttinger liquid.
\end{abstract}

\begin{figure}
\centerline{\epsfig{file=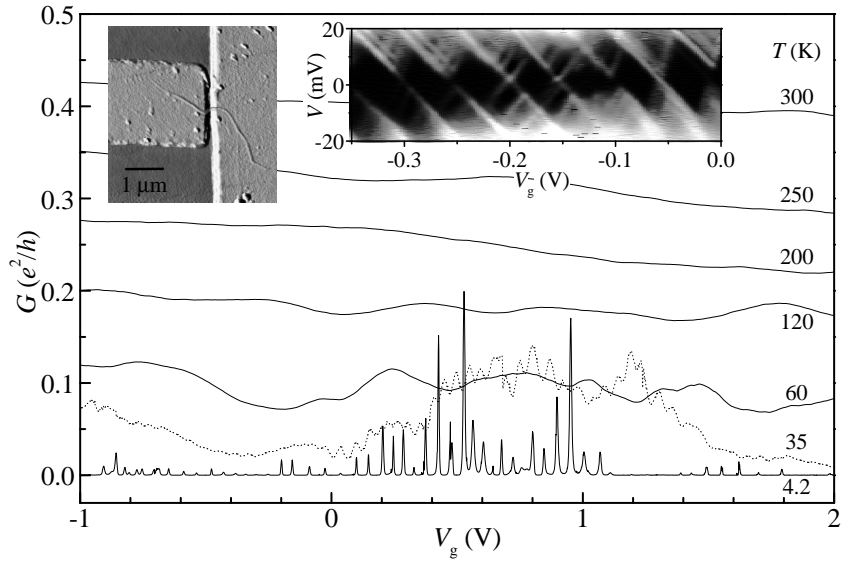,width=12cm}}
\caption{ Conductance $G$ vs gate voltage $V_{\rm g}$ at temperatures $T$ from 300 K to 4.2 K for a end-contacted device.  Right inset: grey scale plot of ${\rm d}I/{\rm d}V$ vs $V_{\rm g}$ and $V$ at 4.2 K (lighter = more positive).  Left inset: AFM image of a nanotube rope lying over two electrodes separated by a 0.2 $\mu$m gap.
}\label{fig1}
\end{figure}

The strength and extended length of single-walled carbon nanotubes makes it quite straightforward to attach metallic electrodes to them.  This has enabled several recent studies of the transport properties of individual tubes and ropes (ordered bundles of tubes) \cite{ref1,ref2,ref3,ref4,ref5}.  In most of these studies the conductance at low temperature $T$ is found to be dominated by Coulomb blockade (CB).  Here we also include measurements on ropes with high conductance and weak $T$ dependences which {\em do not}\ show CB \cite{ref11}.  We analyse the characteristics of all our devices in the light of predictions that electrons in nanotubes should form Luttinger liquids.

Each device consists of an individual nanotube rope \cite{ref6}, containing between 1 and~$\sim 20$ tubes 
lying on a thermally grown SiO$_2$ surface and contacted with gold electrodes patterned by electron beam lithography.  The electrode separation is 0.2 or 0.5 $\mu$m, and the metallically doped silicon substrate beneath the 0.3- or 1.0-$\mu$m thick SiO$_2$ is used as a gate electrode.  We have two varieties of devices: `end-contacted', where the electrode metal is deposited on top of the rope; and `bulk-contacted', where the rope is deposited on top of prefabricated electrodes.  An atomic force microscope (AFM) image of a typical device is inset to Fig.~1.  In all the measurements reported here, the two-terminal dc current-voltage ($I$-$V$) characteristics were measured in a cryostat with the device bathed in helium.

\begin{figure}[t]
\centerline{\epsfig{file=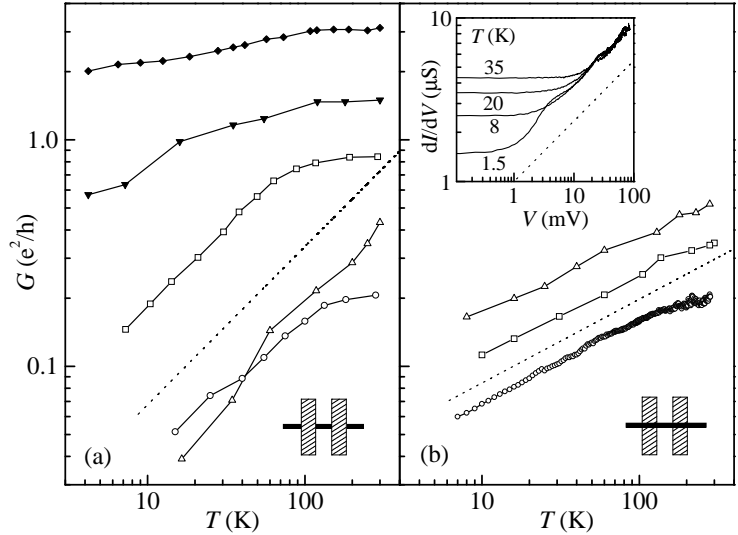,width=12.2cm}}
\caption{ Conductance (averaged over $V_{\rm g}$) vs temperature for several devices.  The data plotted with open symbols have been corrected for Coulomb blockade [3], while the others have not.  (a) End-contacted ropes. The dashed line indicates $G \propto T^{0.7}$.  The open up triangles correspond to the data in Fig.~1.  (b) Bulk-contacted ropes. The dashed line here indicates $G \propto T^{0.37}$. Inset: ${\rm d}I/{\rm d}V$ vs $V$ at several temperatures for a bulk-contacted rope.  The dotted line here indicates ${\rm d}I/{\rm d}V \propto V^{0.37}$.
}\label{fig2}
\end{figure}

Although transistor-like behavior is sometimes seen \cite{ref4}, we consider here only devices whose conductance $G$ is approximately independent of gate voltage $V_{\rm g}$ at room temperature.  
Fig.~1 illustrates the behavior of a typical (end-contacted) device as it is cooled down.  As $T$ decreases, $G$ drops steadily, and below $T \sim 50\ {\rm K}$, Coulomb blockade (CB) peaks develop in $G$ vs $V_{\rm g}$ \cite{ref1,ref2}.  The right inset is a grey scale plot of ${\rm d}I/{\rm d}V$ vs $V_{\rm g}$ and $V$ at 4.2 K.  A standard analysis of this bias spectroscopy plot \cite{ref2,ref8} yields $C_{\rm g}/C\sim 0.15$ and a charging energy $U = e^2/C \sim 10\ {\rm meV}$, where $C_{\rm g}$ is the capacitance to the gate and $C$ is the total capacitance.  The excitation spectrum for each charge state is resolved here at 4.2 K, allowing us to estimate a mean level spacing $\Delta \sim 2$~meV.  From the regularity of this plot we infer that the conductance is largely determined by a single quantum dot. We consider in this paper only devices which show such regularity in their spectroscopy plots.  The values of $U$ and $\Delta$ are consistent with charging a rope segment of length $L$ roughly equal to the contact separation (in this case $L$ = 0.5 $\mu$m).  This is usual for end-contacted ropes.  For bulk-contacted devices on the other hand, $U$ is smaller and corresponds to charging the full length of the rope.  This implies that in both cases the electrode-tube interfaces act as tunnel contacts to the rope which form the dot.  Our terminology was chosen to reflect this: end-contacted means that the current flows from the metal into the end of the active rope segment, and bulk-contacted means it flows into the bulk.

In Fig.~2 the average conductance $G_{\rm av}$ is plotted against $T$ on a log-log scale for several end-contacted (Fig.~2a) and bulk-contacted (Fig.~2b) ropes.  In all cases, $G_{\rm av}$ decreases monotonically as $T$ decreases, irrespective of the room temperature conductance $G_{\rm RT}$, which varies greatly.  That said, we can distinguish the following two categories of behavior.  In the first, which contains most devices, 
$G_{\rm RT} < e^2/h$, and the conductance decreases by up to an order of magnitude before it is 
Coulomb blockaded at low $T$.  For these devices we do not plot $G$ at the lowest temperatures, where CB dominates, and at higher $T$ we multiply $G$ by a correction factor to compensate for the effect of classical CB \cite{ref3}.  Note that this factor is between one and two, and has no major qualitative effect on any of the data. 
In the second category, which contains only end-contacted devices, $G_{\rm RT}> e^2/h$, the dependence on $T$ is much weaker, and there is no Coulomb blockade at 4.2~K.

We begin by discussing the results (open symbols in Fig.~2) for devices in the first category, having $G_{\rm RT} < e^2/h$.  Note that the $T$ dependence for all bulk-contacted ropes in Fig.~2b is very similar, and closely resembles a power law, $G\propto T^{\alpha_{\rm bulk}}$, with $\alpha_{\rm bulk} \approx 0.37$.  For the end-contacted ropes in Fig.~2a the $T$ dependences are steeper, and (at least below $\sim 100$~K) are roughly described by another power law, $G\propto T^{\alpha_{\rm end }}$ with $\alpha_{\rm end}\approx 0.7$.  We have previously argued that these results are consistent with the existence of Luttinger liquids (LL) in nanotubes \cite{ref3,ref9,ref10}.  Tunneling from a metal contact into an LL is expected to be suppressed at low energies, resulting in power laws for the conductance, $G\sim T^\alpha$ (for $eV < k_{\rm B}T$) and 
${\rm d}I/{\rm d}V \sim V^\alpha$ (for $eV > k_{\rm B}T$).  Given a Luttinger parameter of $g\approx 0.28$ for a metallic nanotube \cite{ref3,ref9,ref10}, the exponent is predicted to be $\alpha = (g^{-1}-1)/4 \approx 0.65$ for tunneling into the end of the tube, and
$\alpha = (g^{-1}+g-2)/8 \approx 0.24$ for tunneling into the bulk \cite{ref3}.  These numbers are in fair agreement with the measured quantities $\alpha_{\rm end}$ and 
$\alpha_{\rm bulk}$ above.  Moreover, the prediction that the same exponent $\alpha$ should be seen in the bias as in the $T$ dependence is also borne out.  This is illustrated in the inset to Fig.~2b, where at high bias the ${\rm d}I/{\rm d}V$ vs $V$ traces for a bulk-contacted device converge parallel to a line (on this log-log plot) of slope $\alpha_{\rm bulk} = 0.37$.

\begin{figure}[t]
\centerline{\epsfig{file=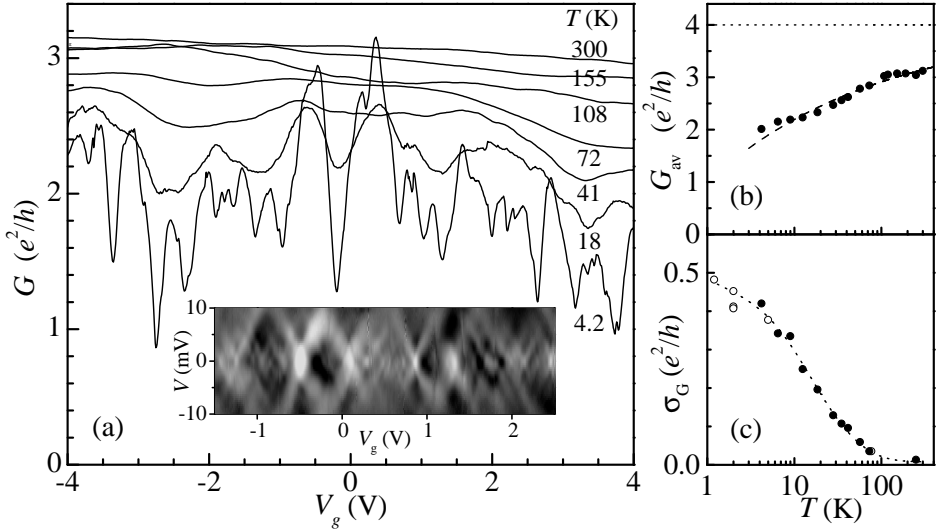,width=13cm}}
\caption{ Gate-voltage characteristics at a series of $T$ for a device with $G_{\rm RT} \sim 3 e^2/h$.  Inset: grey scale plot of ${\rm d}I/{\rm d}V$ vs $V_{\rm g}$ and $V$ at 1.2 K (from a different cool-down).  (b) Mean conductance (for $V_{\rm g}  < 0$) $G_{\rm av}$ vs $T$.  The solid curve is a plot of $G = 4-aT^b$ with $b = -0.22$.  (b) Standard deviation $\sigma_G$ vs $T$.  The solid and open circles are from different cool-downs. The dotted curve is a guide to the eye.
}\label{fig3}
\end{figure}

We come now to the second category of devices, having $G_{\rm RT}> e^2/h$.  Fig.~3 shows the characteristics of a device with $G_{\rm RT} \sim 3 e^2/h$, (diamonds in Fig. 2a), at a series of $T$ \cite{ref11}.  In strong contrast to Fig.~1, there is almost no $T$ dependence above 100 K.  At lower $T$, large, reproducible fluctuations start to develop in $G$ vs $V_{\rm g}$.  In Figs.~3b and 3c we plot the mean conductance $G_{\rm av}$ and the standard deviation $\sigma_G$ of these fluctuations.
The bias spectroscopy plot inset to Fig.~3a exhibits a well behaved, symmetric cross structure.  This implies that the conductance is dominated by transmission through a single object, symmetrically coupled to the two contacts.  The fluctuations are almost unaffected by a magnetic field of 7 T. 
This is consistent with their arising from a nanotube rope, whose thickness is much smaller than the magnetic length (24 nm) at this field.

Nevertheless, these fluctuations are quantitatively and qualitatively different from the CB oscillations in Fig.~1.  We notice that $G_{\rm av}$ remains above $2 e^2/h$, while $G$ fluctuates by as much as $2 e^2/h$, and even occasionally peaks above $G_{\rm RT}$.  These facts are at odds with standard Coulomb blockade.  Rather, we suggest that high-transparency contacts exist between the gold and one or more metallic nanotubes within the rope.  Apart from the high conductance, further evidence for this comes from the very small value of $C_{\rm g}/C \sim 0.01$ deduced from the spectroscopy plot, which means that the rope is much more weakly sensitive to $V_{\rm g}$ than are similar devices that behave as quantum dots.  This is consistent with very strong coupling to the electrodes.  It is therefore possible that the fluctuations are related to transport through the LL in a tube.  Any amount of disorder is expected to completely suppress the conductance of a LL at sufficiently low $T$.  At high $T$ a single impurity produces a power law suppression of $G$ from its ideal value $G_0 = 4 e^2/h$ \cite{ref9}.  In Fig.~3b the dashed line is a fit of $G_{\rm av}$ to the form $G_0 - aT^b$, yielding $b = -0.22$.  Although the fit appears quite good, we make no claim to its validity, as the behavior of a disordered LL remains to be addressed theoretically.

In conclusion, when the conductance of a nanotube rope device is dominated by tunneling from the contacts, power law dependences on temperature and bias are observed which are consistent with the predicted existence of Luttinger liquids in nanotubes.  Further, we find evidence that one can obtain highly transparent contacts between gold and nanotubes, which should soon enable experimental investigations of the intrinsic conducting properties of such a Luttinger liquid.

We would like to thank Poul-Erik Lindelof, Jia Lu, Reinhold Egger, Anders Kristensen and Leon Balents, for help and discussions.

\end{document}